\documentclass[conference]{IEEEtran}
\IEEEoverridecommandlockouts
\usepackage{cite}
\usepackage{amsmath,amssymb,amsfonts}
\usepackage{algorithmic}
\usepackage{graphicx}
\usepackage{textcomp}
\usepackage{xcolor}

\usepackage[colorlinks,citecolor=black,urlcolor=blue,bookmarks=false,hypertexnames=true]{hyperref} 

\graphicspath{ {./images/} }
\def\BibTeX{{\rm B\kern-.05em{\sc i\kern-.025em b}\kern-.08em
   T\kern-.1667em\lower.7ex\hbox{E}\kern-.125emX}}
\begin{document}

\title{Novel prediction methods for virtual drug screening
\\

   \thanks{Identify applicable funding agency here. If none, delete this.}
}

\author{\IEEEauthorblockN{Josip Mesarić}
\IEEEauthorblockA{\textit{Department of Electronic Systems and Information Processing} \\
\textit{Faculty of Electrical Engineering and Computing}\\
      Zagreb, Croatia \\
        josip.mesaric@fer.hr}
}

\maketitle

\begin{abstract}
Drug development is an expensive and time-consuming process where thousands of chemical compounds are being tested in order to find those possessing drug-like properties while being safe and effective. One of key parts of the early drug discovery process has become virtual drug screening - a method used to narrow down search for potential drugs by running computer simulations of drug-target interactions. As these methods are known to demand huge amounts of computational power to get accurate results, prediction models based on machine learning techniques became a popular solution requiring less computational power as well as offering the ability to generate novel chemical structures for further research. Deep learning is to stay in drug discovery but has a long way to go. Only in the past few years with increases in computing power have researchers really started to embrace the potential of neural networks in various stages of the drug discovery process. While prediction methods promise great perspective in the future development of drug discovery they open new questions and challenges that still have to be solved.
\end{abstract}


\section{Introduction}
Virtual screening (VS) is a computational field that has emerged in the past two decades as an aid to conventional experimental drug discovery by using statistics to estimate unknown bio-interaction between compounds and biological targets, usually proteins\cite{wang_pairwise_2014}. VS is a search for compounds with a defined biological activity using a computation model. Molecular compounds are filtered through a progressive series of tests which determine their phyisio-chemical properties, bioactivity, effectiveness and toxicity for later stages of drug discovery process.
“With the emergence of supercomputers, we can analyze in more depth the natural laws of life on Earth, including the metabolic processes of complex diseases like cancer. We can analyze data obtained from high-throughput genomic and proteomic tools, solve systems with millions of linear equations, and analyze graphs that represent thousands of genes and proteins”\cite{tomic_evaluation_2018}.
There are numerous examples where drug discovery systems can be seen as complex adaptive systems with nonlinear emergent behaviour that is hard to predict\cite{schneider_novo_2016} due to its sensitivity to small structural modifications having hugely different effect (“activity cliff”) or unknown interdependencies and gene expression influences. This is inherently also a limitation and challenge for virtual screening methods, along with others that will be mentioned in this review.
Methods of virtual screening use structural models of compounds and target proteins, their physio-chemical properties, bioactivity and experimentally verified bio-interaction data to leverage predictive behavior. 
Although the drug production process still remains expensive and time-consuming due to a high proportion of failures in clinical trials\cite{mullard_2020_2021} (in 2020. FDA didn’t approve 42\% of drugs for rare diseases and 58\% for substantial improvements for serious diseases) some impressive technological leaps were seen during the past few years. Mostly powered by artificial intelligence(AI) and ever-growing omics databases, new computational methods for early phase drug discovery that are developed will be presented in this article. One of key challenges in comparing different AI methods is lack of universal benchmarking datasets by which the scientific community can more precisely compare performance results. Another limitation of deep learning models is that most training/validation sets are still based on handcrafted features or predefined descriptors which makes structural information extraction more difficult from raw inputs. Besides that, most deep model architectures are not initially suited for molecular data representation so researchers have to transform data into more appropriate forms\cite{wan_deep_2016} or develop novel architectures for more accurate molecular representation(e.g. graph neural network\cite{Lau2017BrendanADC}). Deep learning methods have improved different aspects of drug discovery like molecular design, chemical synthesis planning, macromolecular target identification or protein structure prediction\cite{pcm_recent_2017}.

\begin{figure}[htp]
\centering
\includegraphics[width=0.8\linewidth]{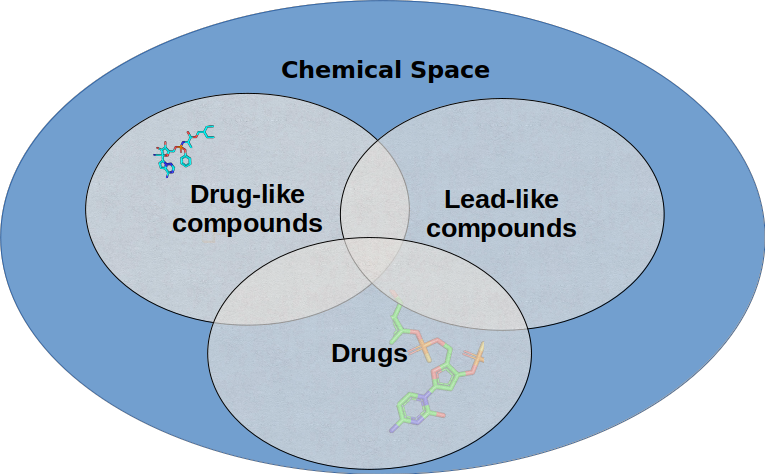}
\caption{Computational prediction of biological activity is still a very challenging task with high rates of false positives and negatives. Some compounds that can be developed into drugs are never identified as drug-like or lead-like compounds. On the other hand, some compounds that have very strong biological activity do not possess drug-like properties, for example they are toxic for humans.}
\end{figure}

Molecules or new chemical entities (NCEs) with low or medium affinity are referred to as “hit” compounds and compounds with sub-optimal target binding affinity are “lead” compounds. Lead compounds are just first iterations of a molecule that is going to be further researched and possibly approved as a drug. To be chosen for further development a lead compound show have following properties: (i) relatively simple chemical features (e.g. suitable for combinatorial/medicinal chemistry optimization, no or few chiral centres), (ii) well-established structure-activity relationship (property where similar compounds or chemical groups share similar biological activity), (iii) favourable absorption, distribution, metabolism, excretion, and toxicity (ADME)\cite{adme_predicting_2002} properties and favourable patent situation for commercialization of discover compounds\cite{sheridan_chemical_2007}.

\section{Virtual screening}
During the hit discovery phase of drug development there are two main approaches: identification of novel bioactive compounds for a target protein and identification of new targets for known bioactive compounds, which are mostly drugs but not exclusively\cite{10.1093/bib/bby061}. Identification and validation of disease-causing genes that are viable as drug targets are key challenges of drug discovery\cite{patel_objective_2013}. On the other side is development of novel compounds that are targeting disease-related protein or repurposing currently approved drugs for treatment of different diseases. 

In the drug discovery pipeline, virtual screening comes before an experimental screening procedure, reducing a set of researched chemicals to only potentially active and drug-like combinations. This is where VS has great potential to reduce the time and cost required for high-throughput screening\cite{schierz_virtual_2009}.
It is a knowledge based method for which prior knowledge about data is necessary. 
Based on used input features VS methods can be grouped into three categories: structure-based, ligand-based and proteochemometric modeling (PCM) \cite{pcm_recent_2017} which just recently advanced so much it can be considered a standalone category. Structure based VS methods use 3D structure of targets and compounds to model target-compound interactions while ligand-based VS methods use molecular properties of compounds, which are mostly non-structural descriptors\cite{10.1093/bib/bby061}. Ligand-based VS is based on the hypothesis that structurally similar compounds have similar biological activities\cite{johnson_concepts_1990}, usually expressed as quantitative structure-activity relationship(QSAR). Its approach consists of finding a compound that is active against some target and using similarity-based techniques to find similar compounds from a library of compounds or by searching chemical space\cite{chemical_space} in case of \textit{de-novo} drug design to identify candidates for further screening process. On the other hand, PCM approach models the interaction by combining non-structural descriptors of both compounds and targets at the input level.

\begin{figure}[htp]
\centering
\includegraphics[width=0.5\linewidth]{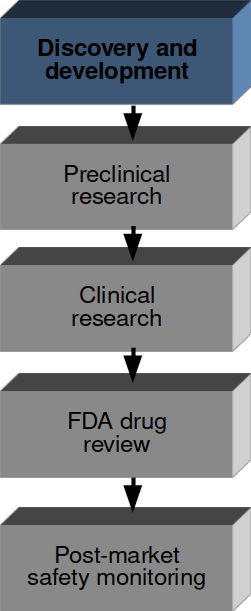}
\caption{Virtual screening is part of the first step in the drug development process. It is a method for low resolution research of vast amounts of compounds from chemical space for “lead” and “hit” chemicals with desired properties. Goal of the drug discovery process can be finding novel drugs (\textit{de-novo}) or repurposing existing drugs for treatments other than currently approved by FDA.}
\end{figure}

\section{Descriptors and features}
Each VS method uses different biological, topological and physio-chemical properties of compounds and targets combined with the experimentally validated bioactivity data to predict biological activities\cite{10.1093/bib/bby061}. For the task of VS it is important to vectorize compounds and targets to get their appropriate computational representation. For quantum mechanical and biophysical datasets, the use of physics-aware featurizations can be more important than the choice of a particular machine learning algorithm\cite{wu_moleculenet:_2018}.
A well defined descriptor should faithfully represent intrinsic physical and chemical properties of the corresponding molecule. This kind of representation enables statistical models and neural networks to learn and generalize the shared properties among the molecules that lead to the interaction between compounds and targets\cite{10.1093/bib/bby061}. Molecular descriptors are numerical vectors representing compound’s features which are generated by algorithms based on geometrical, structural and physio-chemical properties of those molecules.
Descriptors are categorized based on the dimensionality of information describing the molecule, which can get very large as there are more than a thousand different types of molecular descriptors in literature\cite{todeschini_molecular_2009}. 
These extremely high dimensional vectors can be very challenging to work with, creating computational overhead and reducing prediction accuracy. It is known that adding features can improve prediction performance up to a certain point after which it only decreases performance as more features are added to the model. This is known as a \textit{curse of dimensionality} \cite{powell_approximate_2011}. In order to cope with that drawback usually various dimensionality reduction techniques are employed.

3D structure models of proteins and compounds carry important information for determining their bioactivity and function. Currently there is only a small fraction of experimentally known 3D structures of compounds(small molecules) and human proteins, some structures partially and some completely known or predicted. One of the most complete and publicly available databases of such structures is RCSB Protein Data Bank (PDB) \cite{rcsb-pdb}. Most protein structures are predicted from their amino acid sequence with certain accuracy, and some popular methods for protein structure prediction are SWISS-MODEL\cite{waterhouse_swiss-model:_2018} and I-TASSER\cite{yang_i-tasser_2015}. Recently Google’s company DeepMind has released their neural network AlphaFold 2 for which they state that it solves the structure prediction component of the ‘protein folding problem’\cite{dill_protein_2008}. They show that AlphaFold 2 can predict protein structures with atomic accuracy even when there are no known similar structures\cite{jumper_highly_2021}. They have built AlphaFold Protein Structure Database where they provide open access to protein structure prediction of human proteome and some other usually studied oranzims. Currently this is the most complete database of predicted 3D structures of human proteins, containing almost the entire human proteome (98.5\% of proteins found in humans)\cite{tunyasuvunakool_highly_2021}. 
Another project RoseTTAFold inspired by DeepMinds network architecture incorporated a three-track neural network of (i) the 1D sequence level, (ii) the 2D distance map level and (iii) 3D coordinate level information\cite{RoseTTAFold}. RoseTTAFold can also generate accurate protein-protein complex models from sequence information alone, without the need for a traditional docking of two proteins. 
Compound structures are much easier to obtain and most of them are available in public databases like DrugBank\cite{wishart_drugbank:_2006} while the rest is in private databases usually held by pharmaceutical companies. 

Often researchers use already developed tools for other purposes to solve a drug-target interaction (DTI) prediction problem or use a different representation of their input data. For example, using natural language processing (NLP) methods\cite{ozturk_exploring_2020} to represent molecular descriptors like DeepCPI \cite{wan_deepcpi:_2019} where they exploit latent features from large-scale unlabeled compound and protein data by combining feature embedding and deep learning methods. NLP techniques were used to extract useful features of compounds and proteins by representing compounds as “documents” and their basic structures as “words”, whereas protein sequences are represented as “sentences” and all possible three non-overlapping amino acid residues are “words”. After preparing compounds and proteins as language constructs they feed it to multimodal DNN classifiers to make DTI predictions.

\begin{figure}[htp]
\centering
\includegraphics[width=\linewidth]{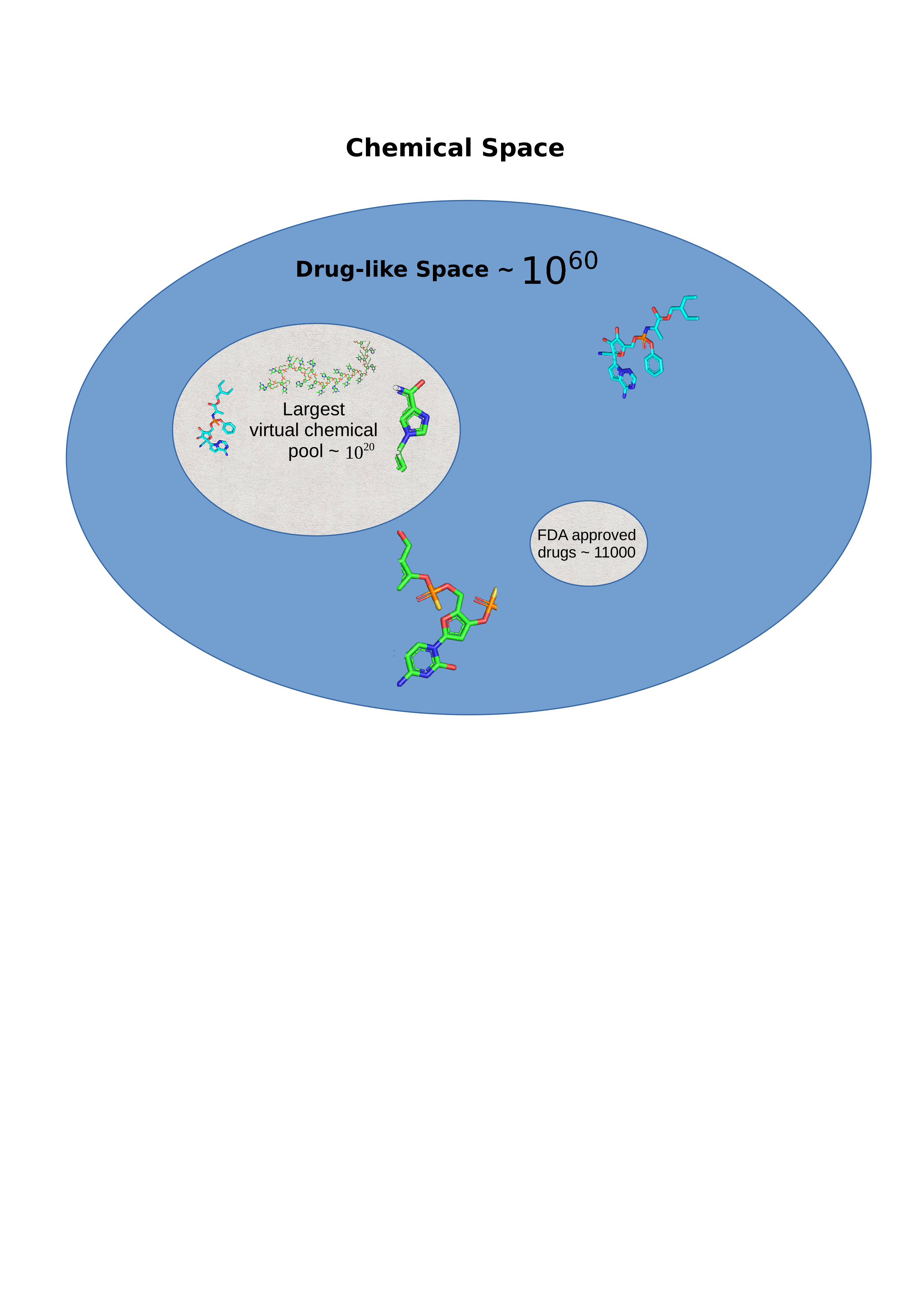}
\caption{Drug-like space is approximated to contain between $10^{23}$ and $10^{60}$. One of the biggest known enumerated chemical libraries GDB-17 contains more than 166 billion compounds. These molecules made of up to 17 atoms are organized in multidimensional space in a way that neighbouring compounds share similar properties, according to 42 characteristics.}
\end{figure}

\section{Drug discovery and design}
\subsection{De-novo drug design}
Method of creating new chemicals with desired drug-like properties with the assistance of computational methods and artificial intelligence. Process of \textit{de-novo} drug design can be formulated as data mining of chemical space, which is a descriptor space of all possible compounds. This process usually starts with a molecular fragment of low molecular weight that is screened against macromolecular targets, which are usually proteins. Molecular fragments can bind at one or more sites on the target protein and that is determined with the help of docking tools like AutoDock Vina\cite{eberhardt_autodock_2021}. Depending on their binding affinity these fragments can be a starting point for development of “lead” compounds. Molecular fragments which are chosen as “lead” compounds usually possess favourable physical, pharmacokinetics and toxicity (ADMET—absorption, distribution, metabolism, excretion, and toxicity) properties.
While developing a molecular fragment into a production-ready drug, or fragment-to-lead optimization, fragment-based methods\cite{de_souza_neto_silico_2020} use strategies like hot spot analysis, druggability prediction, structure-activity relationship (SAR) and applications of various machine learning methods for synthesising new compounds from drug-like fragments. 

Largest \textit{in silico} ligand–receptor docking run through the chemical library has recently been performed in 2019. for 170 million virtual compounds assembled from 130 chemical reactions\cite{lyu_ultra-large_2019}. Just to put that in perspective, the estimated cardinality of a chemical space is in the order of $10^{18}$ - $10^{200}$ molecules\cite{drew_size_2012}, depending on the set of constraints, and drug-like space is estimated between $10^{23}$ and $10^{60}$ molecules\cite{walters_virtual_2019}. Largest publicly available chemical library is ZINC containing ~2 billion compounds\cite{irwin_zinc20free_2020}. There are many proprietary and commercial databases with compound cardinality up to $10^{20}$ and $10^{9}$ respectively. Those libraries mostly contain distinct molecules because intellectual property laws imply that only newly-found molecules can be under commercial ownership. Data-handling on libraries this size is very challenging even in a cloud or HPC parallel environment as memory requirements and search times are increasing with the number of molecules in a library. For example, disk size required to save $10^{20}$ molecules as SMILES is ~200 000 TB in compressed format\cite{van_hilten_virtual_2019}. Techniques aimed at optimizing data-handling in large libraries are beyond the scope of this article and an interested reader can find more information about it from Merck’s Accessible Inventory (MASSIV)\cite{molecules24173096} which is currently one of the largest virtual molecule pools known to the public.

\subsection{Drug repurposing}
Beside discovering new drug candidates, another application of VS is repurposing current drugs for different treatments. Main advantage of screening approved compounds for potential target proteins is that their drug-likeness, bioactivity and non-toxicity are clinically confirmed\cite{ashburn_drug_2004}. Recent approaches to drug repurposing besides conventional chemical similarity and molecular docking use novel approaches in the spirit of systems biology. This means incorporating genomic, proteomic and transcriptomic data in construction of datasets and workflows. Researches now use disease networks which scrape data from various sources to create multilayer networks of known disease interactions (KEGG pathways, text mining)\cite{li_pathway-based_2009}, create Connectivity Map gene expression signatures from mRNA expressions\cite{sirota_discovery_2011} and transcriptomic data\cite{le_transcriptomics-based_2021}.

\section{Machine learning approaches}
Most machine learning approaches can on a high level be described through four steps: data cleaning and preprocessing, feature extraction, model fitting and evaluation of results \cite{angermueller_deep_2016}. Some of those steps are more complex than the others, depending on the approach and problem that is trying to be solved. For example, feature extraction is very challenging in computational biology and chemistry because chemical reactions and atomic structures are hard to describe using numerical vectors. More on features and their numerical encoding can be found in chapter Features and descriptors. 

Authors in \cite{todeschini_molecular_2009} have divided VS methods based on the methodological utilization of the input properties in two groups: similarity-based methods and feature-based methods. Although there are multiple methods in the literature that are using both methodologies this still represents a valid description of general directions for machine learning approaches in VS. 
Similarity-based methods rely on the assumption that biologically, topologically and chemically similar molecules probably have similar bioactivity and biological functions, therefore, they have similar target molecules\cite{ding_similarity-based_2014}. 
This assumption is only partially correct and should be regarded as such because sometimes very small structural changes can have a large effect on bioactivity and properties of those compounds.
Similarity between two compounds is calculated by searching shared molecular substructures and isomorphism. Compound molecules are usually represented as SMILES\cite{smiles_1988} strings and targets are represented as genome sequences where alignment methods are used to calculate similarity. These methods are used to generate similarity matrices which are then used for machine learning models. Similarity methods are especially useful for clustering compounds with similar structure and biological properties. Similarity-based methods are suitable when problems involve heterogeneous data since they can combine different similarity matrices in the same model (e.g. similarity fusion). They are relatively simple to model but require a high number of similarity calculations so they are not computationally always feasible, especially on large datasets. Also, most existing similarity-based methods need to be retrained each time new drugs are approved and they can’t be directly used for novel drug discovery or when adding new targets. There are some novel methods like Weighted k-Nearest Neighbor with Interaction Recovery (WkNNIR)\cite{liu_drug-target_2021} that aim to solve those problems by using ensemble of DTI methods with different sampling strategies and interaction recovery to perform prediction upon a completed interaction matrix without the necessity of retraining and ability to detect previous false negatives.

Feature-based VS methods are based on numerical vector representation of compounds where each compound is mapped to a high dimensional vector which represents different physico-chemical and molecular properties of that compound. Typical workflow of feature-based VS consists of several steps: determining researched targets, finding appropriate compounds from databases of known interactions, generating feature vectors for targets and compounds and finally feeding those vectors to machine learning models to train it for prediction of target interactions. Generated feature vector of unknown compound is used as an input to ML model and output is either active or inactive against targets this predictive model was trained on\cite{westen_proteochemometric_2011}. Feature-based methods are more interpretable than similarity-based methods as they can reveal intrinsic properties of targets and compounds leading, making it easier to identify which features play critical roles in DTIs\cite{10.1093/bib/bby061}. Problem specific features can also be determined (computationally or heuristically) and selected for prediction models to get more accurate results. 

It should be noted that the downside of feature-based methods, and generally most machine learning methods in DTI prediction, is the very difficult construction of a negative training set with negative samples. Databases mentioned in this article contain experimentally validated DTIs which are interacting and they are substantially more numerous than non-interacting pairs. This can lead to a class imbalance problem which results in a bias towards class with most training samples, so DTI prediction models often suffer from a high number of false positive(FP) predictions\cite{10.1093/bib/bby061}. 
Researchers very often create negative training sets by randomly selecting pairs from sets excluding interacting pairs. It is obvious that this method is inherently problematic as some of those pairs can be interacting but their interaction is not yet known. There are some datasets in literature \cite{park_revisiting_2011} of negative example sets for protein-protein interactions where they build two different subsets: (i) unbiased subset for cross-validated testing and predictive performance estimation which should generalize population level and (ii) subset for training predictive algorithms which is constructed to gain best predictive performance of algorithm, even if that subset is biased.

Most ML methods in literature use supervised learning methods and combination of similarity-based and feature-based approaches for constructing predictive models. Such a method \cite{sawada_target-based_repurposing_2015} was employed for drug repurposing by integrating data from multiple sources like ChEMBL database\cite{gaulton_chembl_2017} for compound structures, chemical-protein interactome data and three molecular descriptors: KEGG Chemical Function and Structure (KCF-S) \cite{kotera_kcf-s_2013}, CDK (Daylight-type fingerprints) \cite{steinbeck_cdk_2003} and ECFP4 \cite{rogers_extended-connectivity_2010}. Molecular descriptors were used to construct feature vectors named ‘chemical profile’ and ‘phenotypic profile’ of researched compounds created from manually curated databases like KEGG DISEASE database \cite{kegg_2000}. This is usually the case when researchers are creating databases for training predictive models, they have to merge knowledge from programirably accessible databases containing huge amounts of data with data available in literature.

\subsection{Network-based methods}
Another highly studied topic regarding methodological approaches used in modeling target-compound pairs are network based drug-target interaction prediction methods. Those methods are also referred to as graph based DTI methods because they represent target-compound pairs as graph nodes with edges connecting them. This approach borrows various tools and methods from graph theory, social networks, and biological systems which are employed for link prediction tasks. Main advantage of the network-based approach is the low number of necessary examples for training, which is especially useful in drug design where example space is often very sparse. 

\begin{figure}[htp]
\centering
\includegraphics[width=\linewidth]{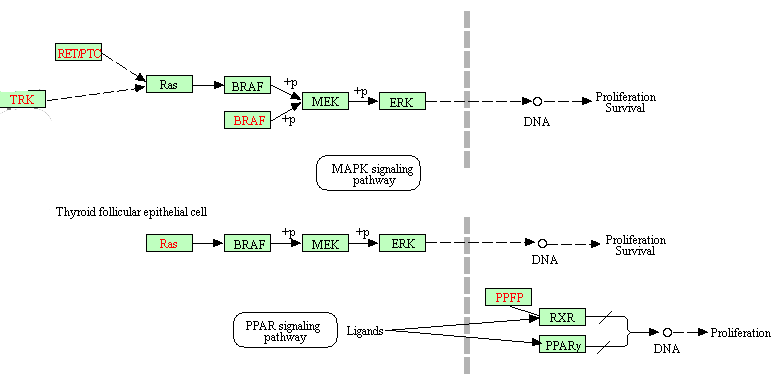}
\caption{Example of KEGG signaling pathway for thyroid cancer in humans. This signaling pathway is a network representation of important elements (e.g. proteins, enzymes, molecular interactions and reactions) for thyroid cancer curated from the literature. Boxes represent ortholog groups, circles are other molecules, usually chemical compounds, and lines represent reactions in metabolic maps.}
\end{figure}

Network-based approaches are also very useful for integrating heterogeneous biological data from different sources for drug-target interaction prediction. Yamanishi et al. \cite{yamanishi_drug-target_2010} have joined chemical space\cite{chemical_space}, pharmacological space\cite{pharmacological_space} and topology of drug-target interaction networks in a unified prediction framework. They showed that DTIs are more correlated with pharmacological effect similarity than with chemical structure similarity. This was the first method to be able to predict potential pharmacological similarity for any drug candidate in a unified way by integrating chemical, genomic and pharmacological data and combining it with known DTIs.

Another example of research that is utilizing biological networks for DTI is an \textit{in silico} model of cancer called VINI. Tomić et al. \cite{vini} have developed a virtual screening framework which transforms the metabolic pathways of cancer from Kyoto Encyclopedia of Genes and Genomes\cite{kegg_2000} into the binding energy matrices. VINI is able to search libraries containing 3D structures and FASTA drug sequences of small molecules and protein-based drugs, to identify those with the highest binding energy to target molecules. The VINI model uses Autodock Vina software in the process of calculating binding free energy ($\Delta$G) between receptors and small molecules, and molecular dynamics tools to calculate $\Delta$G between receptors and protein-based drugs. By definition, $\Delta$G has always negative values expressed in kcal/mol units, or zero value in case there is no binding between receptor and ligand. The lower are these negative values, the greater the inhibition of the receptor by ligand is. Although receptors can be any kind of chemical compounds, most common receptors in KEGG pathways are proteins. Unlike receptors, most researched drug candidates are usually small molecules composed of several to several dozen atoms with low molecular weight. In literature only a few protein drugs (1000-2000 atoms) were screened against KEGG pathways due to the extremely high computational resources demand of molecular dynamic tools (e.g. GROMACS \cite{gromacs_2015}) needed to perform \textit{in silico} protein-protein docking and calculation of their binding free energies and high complexity of producing drugs of such complicated molecular configuration. 

Another example is the computational method DDR \cite{DDR:graph_mining} where authors use a heterogeneous graph containing known DTIs with multiple similarities between drugs and between target proteins. They integrate a heuristic optimization process to obtain the best subset of similarity combinations for a nonlinear similarity fusion method. Similarity fusion is a process of combining selected similarity measures into one final composite similarity that contains information from its components. 

As already stated, network-based models are very suitable for integrating heterogeneous data from different network models, sequencing data and various proteomic analysis into a single model capable of significantly improving personalized and precision medicine\cite{zhou_network-based_2020}. Those models are able to more faithfully represent a single patient's biological environment and predict drug efficiency for that patient with significantly improved accuracy. There is an increasing gap between what is technically possible and what is allowed by privacy laws\cite{price_privacy_2019}. Labeled medical data of patients is subject to very strict privacy legislation so prediction systems should include confidentiality by design. One example of such a system is Swarm Learning\cite{swarm_2021} where authors are uniting decentralized machine-learning, edge computing and blockchain-based-peer-to-peer networking for detection of heterogeneous illness while maintaining confidentiality with the need for a central coordinator.

\subsection{Protein-to-protein binding affinity prediction}
Central element of almost all cellular processes are protein-to-protein interactions(PPI), so knowledge of the structure of protein-protein complexes and predicting their binding affinity is of major importance to understand PPI and its impact on the virtual screening process \cite{computational_prediction_ppi_2020}. Protein-to-protein docking and binding affinity prediction are substantially more complex and computationally demanding than those for protein-to-small-molecule (which most target-compound pairs in literature usually are). There are some novel rigid body protein-protein docking methods that provide less accurate, yet acceptable models for more complexes than flexible docking methods. Authors in \cite{ppi_docking_performance_2020} are archiving protein-protein docking by fast Fourier transform algorithm used for sampling billions of complex conformations. 
There are multiple available scoring schemes and conformational sampling for protein-protein complex structure prediction, while binding affinity approaches can be grouped into structural and sequence based approaches\cite{GROMIHA201731}. There are several software tools\cite{amino_acid_ppi}\cite{ofran_isis:_2007}\cite{vangone_contacts-based_2015} for protein interactions and structure geometry analysis of protein-protein complexes like PSAIA\cite{PSAIA} which uses random forests or web server HDOCK\cite{yan_hdock:_2017} for DNA/RNA docking based on a hybrid model of template-based modeling and free docking. Authors of web based service PSOPIA\cite{murakami_homology-based_2014} for homology-based PPI prediction use combination of three sequence based features: sequence similarities to a known interacting protein pair, statistical propensities of observed interacting protein pairs and sum of edge weights along the shortest path between homologous proteins in a PPI network. There are also studies that focus specifically on some subgroups of proteins to gain better prediction accuracy, like InterPep\cite{protein_peptide_predicting_2019} which is a tool for predicting protein-peptide interactions. 

\section{Deep learning methods}
Deep learning (DL) is a computational model composed of multiple processing layers capable of learning data representations with multiple levels of abstraction. Term “deep” defines a subset of machine learning methods that use multiple hidden neural network layers in their architecture layers\cite{Wen20171401}. DL models can detect intricate relationships and structure in large datasets by using the backpropagation algorithm\cite{10.1093/bib/bby061}. Deep neural networks (DNNs) is an artificial computational architecture inspired by networks of neurons formed in the brain, a multilayer network that is represented as a graph with nodes representing neurons and edges representing neural connections.
Basic example of feedforward DNN is multilayer perceptron - MLP but there are many other implementations and variations in literature. For example, deep convolutional nets and deep recurrent nets are improving state-of-the-art in speech and visual objects recognition, text analysis and also biological domains like drug discovery and genomics \cite{lecun_deep_2015}.

Most commonly used machine learning techniques for DTI prediction in literature past few years were variations of DNNs with multiple hidden layers \cite{ma_deep_2015} because of their ability to make better prospective predictions than random forest (RF)\cite{random_forest_bias_2007} or support vector machine (SVM)\cite{svm_cancer} which were commonly used methods in the early 2000s. This is actually a revival of neural networks because they wever superseded by RF and SVM as more robust back at that time. 
Each layer of DNN captures specific features from multidimensional, often heterogeneous, data. They are capable of detecting higher-level abstractions from complex sets of features and identifying unknown structures and relationships from raw data in a hierarchical manner. 
Convolutional neural networks (CNNs) have gone a long way in image classification\cite{mask_r_cnn} so they are often used in VS when input data can be represented as an image or object with image-like features. In the next chapters of this review, deep learning models are categorized by the neural network architecture they implement.

\subsection{Feedforward deep neural networks}
Feedforward deep neural networks (FF-DNNs) are the first and simplest type of neural networks. Their nodes never form a cycle so information moves in only one direction, forward from input nodes, through hidden layer(s) to output nodes. Architectures like a single or multi layer perceptron are most common although there are architectures implementing a directed acyclic graph which means some input can skip hidden layers.
Authors in \cite{stokes_deep_2020} have used feedforward DNN to discover a molecule from the Drug Repurposing Hub\cite{corsello_drug_2017} that is structurally as different as possible from conventional antibiotics while keeping the same biological activity so it can be used against a wide spectrum of resistant bacteria. Variation of FF-DNN called message passing neural network (MPNN)\cite{yang_analyzing_2019} was used to predict quantum properties of organic molecules which are otherwise computationally expensive to calculate because of a density functional theory (DFT)\cite{becke_densityfunctional_1993} modeling. Three different optimizations were used on MPNN architecture: adding additional molecule-level features from RDKit for large molecules, ensembling with different initial weights and hyperparameter optimization. This speed-up allowed authors to screen multiple chemical libraries in search for structurally distant molecules from known antibiotics.

\subsection{Pairwise input neural networks}
Pairwise input neural networks (PINNs) are a variation of neural networks where the input vector is a combination of both target and compound description vectors. There are some PINNs where a combination of input vectors is processed by the same connected neurons while others use separate groups of neurons for processing target and compound input vectors before they are merged together in a fully connected layer. Pairwise representation is inherently convenient for prediction of pairwise relations like DTI\cite{10.1093/bib/bby061}. Authors in \cite{wan_deep_2016} used unsupervised learning to extract low-dimensional representation of input features which they embedded to a fixed low dimensional space. Amino acid triplets in proteins and compounds molecular structure were represented as word-embeddings so natural language processing (NLP) techniques could be applied to extract hidden structures and relationships. 

\subsection{Convolutional neural networks}
Convolutional neural network (CNN) is a neural network modelling approach inspired by neuroscience to imitate images within the visual cortex. In the visual cortex neurons respond to stimuli in their receptive fields, with some neurons overlapping their parts of their receptive fields. CNN architecture is especially adapted to multidimensional input obtained from image pixels, where each hidden layer is recognizing certain critical features such as lines or objects. CNNs ability to to exploit spatial or temporal correlation in data makes CNN techniques very suitable for usage in computer vision and voice recognition\cite{khan_survey_2020}. Facebook’s Mask R-CNN has outperformed all single-model entries on every task in image instance segmentation, bounding-box object detection and person keypoint detection\cite{mask_r_cnn}. Several studies have employed already developed image recognition CNN models for the purpose of virtual screening, biology activity or toxicity predictions, representing input data as images. Authors in \cite{goh_how_2018} represented each compound as an 80x80 pixel image based on their 2D drawing from chemical databases. These images coupled with their biological properties were fed as input to CNN for classification. Model was used for prediction of toxicity, biological activity and solvation free energy. In other research, authors represented drug molecules as 2D graphs with atom features and fed it to CNN to learn molecular features from molecular graphs. These molecular features were used as a dataset for training other prediction models\cite{Altae-Tran2017283}. Their motivation was to significantly lower the amount of data required to make meaningful predictions in drug discovery applications. It was implemented as one-shot learning architecture combining iterative refinement long short-term memory and graph convolutional neural network, which is described in more detail in a separate section. Besides the need for large amounts of training data CNN’s main drawback is its inability to model potential relationships between distant atoms in a raw molecule sequence. Authors in \cite{BECK2020784} have used a combination of CNN and self-attention mechanism\cite{shin_self-attention_2019} (concept initially developed for natural language representation learning) to overcome short backs of CNN for prediction of commercially available antiviral drugs that could act on SARS-CoV-2 virus.
In another example, authors used atom and amino acid embeddings in creating vector representation of protein-ligand complexes further processed by CNN\cite{pereira_boosting_2016}. Here they are using CNN to boost the docking process for virtual screening, which is usually performed by docking software like AutoDock Vina or Dock.

\subsection{Recurrent neural networks}
Recurrent neural networks (RNN) is a type of deep learning architecture designed specifically to handle sequenced data which has achieved great success in fields like NLP and voice recognition. Main difference between FF-DNNs and RNN is that in deep learning methods that follow feedforward architecture there is no connection between nodes in the same hidden layer, only between nodes of different adjacent layers. In RNN architectures there is also a connection between nodes in the same hidden layer which enables them to process sequential information by affiliating adjacent hidden nodes with each other and capturing the calculated information from previous time slices and storing it for the subsequent processing. 
Authors in \cite{gupta_generative_2018} used RNN containing long short-term memory (LSTM) cells for de novo drug design. It captured SMILES string molecular representation and learned probability patterns that were used for generation of new SMILES strings. By employing transfer learning this RNN architecture can be fine-tuned for specific molecular targets. In transfer learning the model is keeping information from the previous target while tasked to solve a different but related and yet unseen target. Authors in \cite{olivecrona_molecular_2017} are using RNN to generate novel molecules with some desirable properties, fine-tuning it through augmented episodic likelihood. NN was tested through three step assignment: (i) generating sulfur-free molecules, (ii) generating analogues of the drug Celecoxib and finally (iii) generating novel compounds against an assigned target with 95\% of them predicted to be active.

\subsection{Restricted Boltzmann machine (RBM)/Deep belief networks (DBN)}
Restricted Boltzmann machine is a variation of the single-layer model Boltzmann machine\cite{bengio_learning_2009} in which neurons from visible and hidden layers must form a bipartite graph. “Restricted” implies that they have symmetric connection between them but there are no connections between nodes inside the same layer which allows usage of more efficient training algorithms like gradient-based contrastive divergence algorithm or likelihood gradient \cite{montavon_practical_2012}. By “stacking” RBMs into a deep belief network(DBN) they can be used in deep learning, where usually gradient descent and backpropagation are used for fine-tuning the results. 
Authors in \cite{wang_predicting_2013} trained individual RMB for each target and integrated them together into a final model. Main goal of this architecture was to incorporate a heterogeneous set of compounds and targets with different DTIs into a single multidimensional prediction model. They divided interactions as direct and indirect, where physical binding of a compound molecule to a protein is considered direct interaction and all other interactions are considered indirect, for example changing expression levels of genes responsible for encoding target proteins. 
On the other hand, authors in \cite{Wen20171401} grouped all target proteins in the same group to train a single predictive model. They only later after training decoupled their model to test the ability of DBN in abstracting of the input data and generating more informative representation in each succeeding hidden layer. They used each layer’s representation to train a simple logistics regression model to predict DTIs. Accuracy of the logistic regression model was getting higher as the hidden layer used for training was deeper in DBNs architecture. This indicated that each hidden layer in DBN was generating more informative and meaningful representation of input data.

\subsection{Graph convolutional neural network}
Graph convolutional networks (GCN) achieved remarkable success during the past 10 years in various domains of bioinformatics. They were developed through efforts to generalize a convolution operator on non-Euclidean structured data, which are in this case graphs. Authors in \cite{sun_graph_2020} point out that graph-based representation of molecules has several advantages over descriptor-based molecule representations: it extracts features with respect to structure of the data and it can automatically extract features from raw input instead of hand picking features and thus potentially leaving out important information or creating bias. 
GCN found their applications in molecular property and activity prediction, interaction prediction, synthesis prediction and \textit{de-novo} drug design\cite{wang_pairwise_2014}. Two main approaches to building GCN for virtual drug screening are distincted by their definition of convolution: spatial and spectral GCN. Spatial GCN formulates convolution in spatial domain by summing up all features vectors of neighborhood nodes in a graph, while spectral defines convolution in the graph spectral domain\cite{Defferrard20163844}. Spatial convolution constructs convolution explicitly in the vertex domain. Representation of a single node is updated by recursively aggregating information from its neighborhood nodes. Spectral convolution takes variation of Fourier transformation as the basis for signals defined on a graph, typically focusing on the properties of the graph Laplacian matrix. 
Authors in \cite{YingkaiGao20183371},\cite{Duvenaud20152224} used GCNs to overcome common challenges of traditional machine learning used for DTI prediction which are: (i) extensively large chemical space on which search for drug-like compounds is performed, (ii) interaction prediction is usually defined as binary classification problem (e.g. compound is either active or inactive for a single target) making it ineffective for screening against new targets and (iii) lack of biological interpretation.

\subsection{Generative deep neural networks}
Besides processing labeled data in supervised learning, DNNs can be used for analyzing non labeled data in an unsupervised learning process. This is usually achieved with two NNs (e.g. DBN) which act like encoder and decoder, where former maximizes its classification accuracy and latter is maximizing its discrimination accuracy. 
Most often seen generative neural networks in literature are variational autoencoder (VAE) and adversarial autoencoder (AAE). Main difference between these two approaches is in the way they compute loss in latent representations. VAE relies on Kullback–Leibler divergence\cite{kullback_information_1951} between distribution obtained from the encoder and prior distribution of input data - we would like to get latent distribution as close to gaussian prior as possible. On the other hand, adversarial AE uses a different approach for getting the term minimized adversarially. AAE achieves the same result by sending samples from latent (“fake”) distribution and samples from a gaussian distribution (“real” samples) to a discriminator which should learn to produce samples similar to prior gaussian distribution.
Authors in \cite{kadurin_drugan:_2017} have used deep generative adversarial network (GAN)\cite{creswell_generative_2018} called druGAN to demonstrate a proof-of-concept for adversarial autoencoder (AAE) used to identify new molecular fingerprints with predefined anticancer and drug-like properties. They show improvement over variational autoencoder (VAE) in terms of adjustability and capacity of processing large molecular sets. Autoencoders have been commonly used for molecular feature extraction purposes while building prediction models for DTI. Various types of autoencoders have capabilities of extracting important features from high dimensional feature space describing targets and compounds, allowing researchers to get better results with sparse datasets\cite{Alain2013RegularizedAE}. There are many examples in recent literature\cite{polykovskiy_entangled_2018}\cite{kadurin_cornucopia_2017} of researchers using autoencoders to extract favourable features from molecular descriptors which they later use for synthesis of novel compounds with the same features through generative adversarial networks.
Authors in \cite{Kadurin201710883} used generative adversarial autoencoders (AAE) for generating novel molecular fingerprints with a defined set of parameters. NCI-60 cell line was used to train AAE with input and output defined as a combination of binary fingerprint vector and log concentration of the molecule. Additional neuron was added to control growth inhibition (GI) percentage which indicates reduction in the number of tumor cells if its value is negative. AEE was trained on MCF-7 cell line assay data for 6252 compounds based on their log concentration, molecular fingerprint and GI data. AEE and decoder were used to generate 32 descriptor vectors containing molecular fingerprint, GI value and concentration. This set of vectors was later screened against 72 million compounds from Pubchem\cite{pubchem} where top 10 hits with maximum likelihood function were filtered. Pubchem BioAssay database \cite{pubchem_bioassay} was used to identify compounds with biological and medical relevance and it was discovered that trained AEE generated some very similar compounds to those that are known to have anticancer activities, mostly similar to anthracyclines.

\subsection{Explainable AI in drug discovery}
DNNs are often limited in their usability due to their black-box nature so different model interpretations were developed for better understanding of the inner-process of training NN and their prediction results\cite{10.1093/bib/bbaa177}. 

Explainable, or sometimes referred to as interpretable, artificial intelligence is a collection of methods developed to gain better insight in the training and prediction process of deep learning algorithms. There are several properties of explainable artificial intelligence that are favourable in drug discovery process: (i) transparency (knowing how certain result was reached, e.g. which features play major role), (ii) justification (reasoning why that answer was provided by the model), (iii) informativeness (ability to provide additional, previously unknown, information for human operators) and (iv) uncertainty estimation (quantified measure of how reliable is models prediction)\cite{Lipton2018}. There are many domain-specific challenges in building explainable AI for drug discovery - first of all, drug discovery has inherently problematic modeling of raw data. Unlike image recognition or natural language processing, molecules don’t have complete and raw data representation. It is still not clear which is the best molecule representation and often it is very dependent on the domain of application. This choice of molecular representation becomes a limiting factor in performance and explainability of the model. Molecular representation determines which properties can be explained and which chemical information can be retained (for example, if we represent a molecule with its physicochemical properties then the model can only be interpretable in a regard to those specific properties)\cite{jimenez-luna_drug_2020}.

\section{Conclusion and future challenges}
Appeal of applying artificial intelligence in the drug discovery process is in its potential to build data-driven models from vast datasets generated from high throughput screening(where hit rates are only 0-0-1\%) and to prioritize alternative variations. In the beginning this implies synergy of machine and human intelligence in a specific domain where machines augment capabilities of medicinal chemists in early phases of drug design and selection. Ultimately, the goal is to develop a model capable of autonomously generating new chemical entities (NCEs) of desired properties without the need for human intervention and expensive high throughput screenings. From Schneiders et al.\cite{schneider_rethinking_2020} perspective AI-aided drug design should provide questions to five ‘grand challenges’ in order to be successful in its mission. Those grand challenges are: “obtaining appropriate datasets, generating new hypotheses, optimizing in a multi-objective manner, reducing cycle times and changing the research culture and creating appropriate mindset”. Authors point out that input dataset are key factors when evaluating predictive models and whether those dataset were collected with the final goal in mind. In most cases the relationship between \textit{in vitro} and \textit{in vivo} tests for animal and human patients is not clearly established so models built on surrogate data have inherent limitations. This extends to designing high-throughput screening datasets, protein structure prediction and appropriate meta-data and annotations in databases. There is a lot of work ahead for researchers, with high necessity for interdisciplinary collaboration in order to harvest the full potential of new predictive methods in drug discovery.

\bibliographystyle{ieeetr}
\bibliography{references}
\end{document}